# Laser Annealed SiO$_2$/Si$_{1-x}$Ge$_x$ Scaffolds for Nanoscaled Devices, Synergy of Experiment and Computation

Damiano Ricciarelli*,[a] Jonas Müller,[b] Guilhem Larrieu,[b,*] Ioannis Deretzis,[a] Gaetano Calogero,[a] Enrico Martello,[a] Giuseppe Fisicaro,[a] Jean-Michel Hartmann,[c] Sébastien Kerdilès,[c] Mathieu Opprecht,[c] Antonio Massimiliano Mio,[a] Richard Daubriac,[b] Fuccio Cristiano,[b,‡] Antonino La Magna[a,*]

[a] *National Research Council, Institute for Microelectronics and Microsystems (IMM-CNR), VIII Strada Catania, 95121, Italy*
[b] *The French National Center for Scientific Research, Laboratory for Analysis and Architecture of Systems (LAAS-CNRS) and University of Toulouse, 7 Av. du Col. Roche, Toulouse, 31400, France*
[c] *Université Grenoble Alpes, CEA-LETI, Minatec Campus, 17 rue des martyrs, Grenoble Cedex 9, F-38054, France*
[‡] *Deceased, January 2024*

**Email**: Reach out to us via institutional websites.

## ABSTRACT

Ultraviolet nanosecond laser annealing (UV-NLA) proves to be an important technique, particularly when tightly controlled heating and melting are necessary. In the realm of semiconductor technologies, the significance of nanosecond laser annealing (NLA) grows in tandem with the escalating intricacy of integration schemes in nano-scaled devices. Silicon-germanium alloys have been studied for decades for their compatibility with silicon devices. Indeed, they enable the manipulation of properties like strain, carrier mobilities and bandgap. In this framework, they can for instance boost the performances of p-type MOSFETs but also enable near infra-red absorption and emission for applications in photo-detection and photonics. Laser melting on such type of layers, however results, up to now, in the development of extended defects and poor control over layer morphology and homogeneity. In our study, we investigate the laser melting of ~700 nm thick relaxed silicon-germanium samples coated with SiO$_2$ nano-arrays, observing the resulting material to maintain an unaltered lattice. We found the geometrical parameters of the silicon oxide having an impact on the thermal budget samples see, influencing melt threshold, melt depth and germanium distribution.

**Introduction**

Integrating ultraviolet nanosecond laser annealing (UV-NLA) with pulsed power emission (pulse duration below $10^{-6}$ s) into thermal processes for micro- and nano-electronics provides adaptable and powerful solutions within extremely limited space and time constraints. The laser-induced heat effectively melts doped semiconductor layers. In the ensuing re-crystallization phase, dopants move from interstitial to substitutional sites, resulting in electrical activation. Moreover, the rapid solidification of melted material prevents the formation of disordered or amorphous semiconductor domains. The use of a laser wavelength in the ultraviolet range results in a melting of well-defined regions at the nanoscale with the advantages of a better control of the involved junctions, avoiding possible damage of neighboring parts of the device. The even redistribution of dopant atoms is facilitated by the high diffusivity in the liquid phase (e.g. $10^{-4}$ cm$^2$ s for liquid Si). Additionally, non-equilibrium segregation during rapid solidification enhances the incorporation of dopants into the lattice. These distinctive characteristics make laser annealing a widely adopted post-fabrication annealing technique in microelectronics. [1-10]

Silicon germanium alloys, $Si_{1-x}Ge_x$, have attracted much interest for decades in microelectronics and other related domains. The higher hole mobility, smaller band-gap and relatively small lattice parameter mismatch between $Si_{1-x}Ge_x$ and Si can boost the performances of p-type MOSFETs, but also enable an optical response in the near infra-red regime, suitable for applications in photo-detection or photonics.[11-21] Additionally, these alloys exhibit interesting physical properties that have some impact during laser annealing. For example, the coexistence of Si and Ge in the lattice structure hinders phononic transport, resulting in a U-shaped thermal conductivity curve as a function of the alloy fraction coordinate (x) and the optical properties are also modified as x changes, mainly due to the alteration of materials' plasmons and phonons.[22-26] Subjecting pure silicon-germanium layers on Si substrates to laser irradiation results in a scarcely controlled morphology, with the appearance of stacking faults and lateral Ge segregation regions due to the formation of a quasi-cellular liquid/solid interface.[27-31]

Capping silicon with anti-reflective silicon dioxide stripes was shown by numerous authors to be valuable in controlling the laser irradiation of samples.[32-36] To the best of our knowledge, the study of $SiO_2$ atop $Si_{1-x}Ge_x$ layers has not been explored. In this paper, we have investigated the interest of such a strategy on ~700 nm thick relaxed silicon-germanium layers with 50 % Ge content, labelled as r-$Si_{0.5}Ge_{0.5}$, capped with nanometer size arrays of $SiO_2$ stripes created with e-beam lithography. The strategy aims to hinder the formation of the quasi-cellular liquid/solid interface and to suppress any possible random nucleation of the liquid. The more confined heating induced by the stripes seeks to reduce final Ge inhomogeneity and stacking faults of the treated samples.[30] Specifically, we have examined laser-irradiated structures with high-angle annular dark-field scanning transmission electron microscopy (HAADF-STEM) and electron energy loss spectroscopy (EELS). We have also used modeling to comprehensively characterize process outcomes. The incorporation of such nano-arrays of stripes turns out to be a promising solution to enhance process control, with the addressing of aforementioned morphological and thermal issues.

**Results and Discussion**

Samples with arrays of $SiO_2$ nanometer-scaled stripes on relaxed silicon-germanium with an alloy fraction of 0.5, labelled as r-$Si_{0.5}Ge_{0.5}$, are schematically shown in Figure 1. Figure 1a schematized sample include several arrays of lines with different widths and pitches. It has a 10 x 10 $\mu m^2$ surface and is made of three distinct layers: the surface layer containing the spaced silicon oxide lines, a middle SiGe stack made of a 700 nm thick $Si_{0.5}Ge_{0.5}$ layer on top of a 4.7 μm thick linearly graded $Si_{1-x}Ge_x$ buffer and, lastly, a bottom Si substrate. Geometrical and morphological features of the samples can be described with the simplified two-dimensional xy section shown in Figure 1b. The two-dimensional representation considers the periodicity along the x-axis (transverse to the $SiO_2$ arrays) and the uniformity of the sample along the z-axis (longitudinal to the arrays). The width of uncapped r-$Si_{0.5}Ge_{0.5}$ between $SiO_2$ stripes is equal to the difference between pitches (P) and widths (W). The height of $SiO_2$ stripes along the y-axis is the H parameter. Arrays of $SiO_2$ stripes on r-$Si_{0.5}Ge_{0.5}$ were fabricated by electron-beam nanolithography (EBL, refer to the methodological section for additional details) with the combination of the following width/pitch values: 30, 50, 70, 80 nm for W and 100, 120, 140, 160, 180, 200 nm for P. Meanwhile, H was fixed to 50 nm. Scanning electron microscopy (SEM) images of fabricated structures prior to NLA process are shown in Figure 2a. When the spacing between $SiO_2$ stripes was low (i.e. e.g. the difference between P and W is small), a coalescence of the HSQ resist during EBL occurred. For higher spacings, well-defined and shaped nano-stripes were obtained. This was expected, as EBL resolution was demonstrated to be limited to 60 nm by enhanced proximity and charging effects, because of the high Ge content of the template. This also resulted in a size increase of patterns.[37] Let us focus on individual selected W/P combinations (e.g. on Figures 2b-g). Stripes are distinct for 30/200, 30/160, and 80/200 W/P combinations, i.e. for nominal spacings between stripes >90 nm. For 30/100, 80/100, and 80/160 W/P combinations with nominal spacings <90 nm, they begin to coalesce.

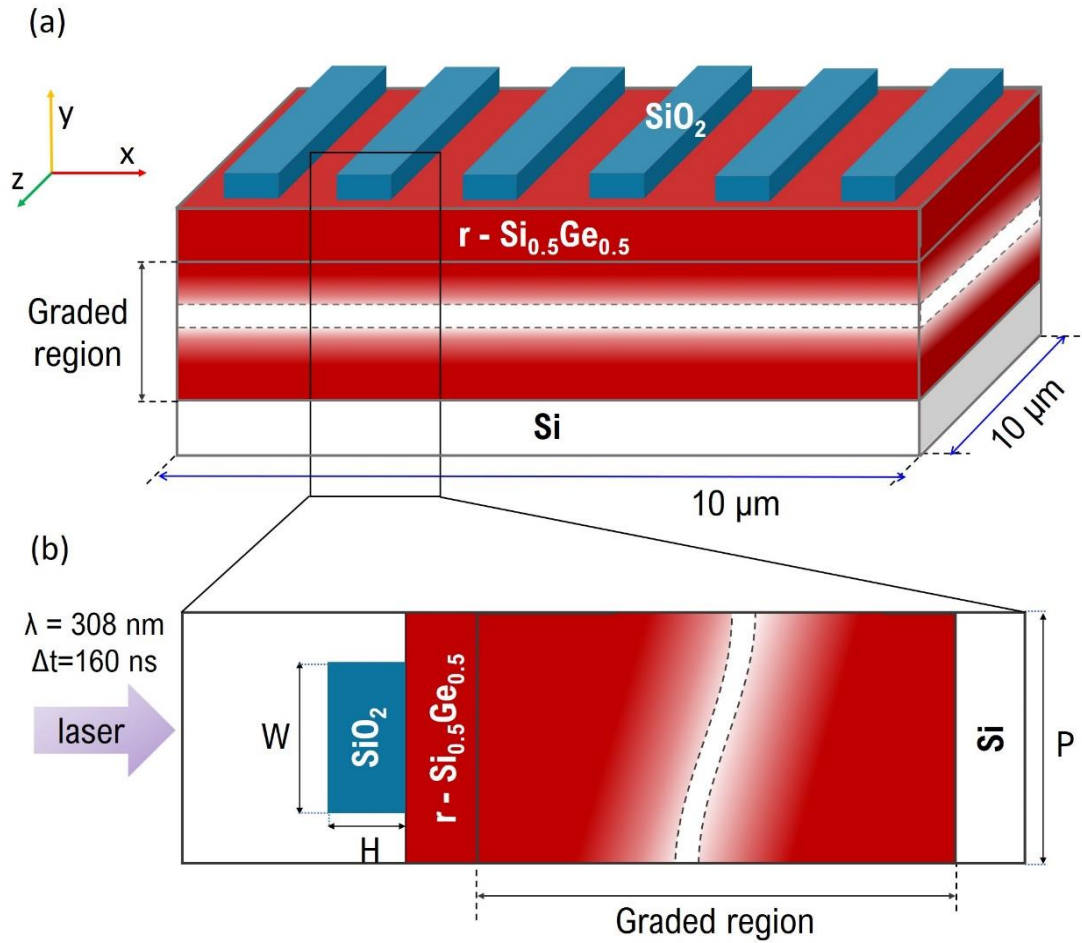

**Figure 1.** (a) Simplified scheme of manufactured samples with arrays of $SiO_2$ stripes on r-$Si_{0.5}Ge_{0.5}$, (b) schematics of the two-dimensional approximation used to classify and model the aforementioned samples. $SiO_2$ stripes' width, pitch and height are symbolized by W, P and H respectively. Periodic boundary conditions are set along the x-axis. The $Si_{1-x}Ge_x$ region with the white break represents the graded buffer layer where the Ge fraction linearly decreases from 0.5 down to 0.0.

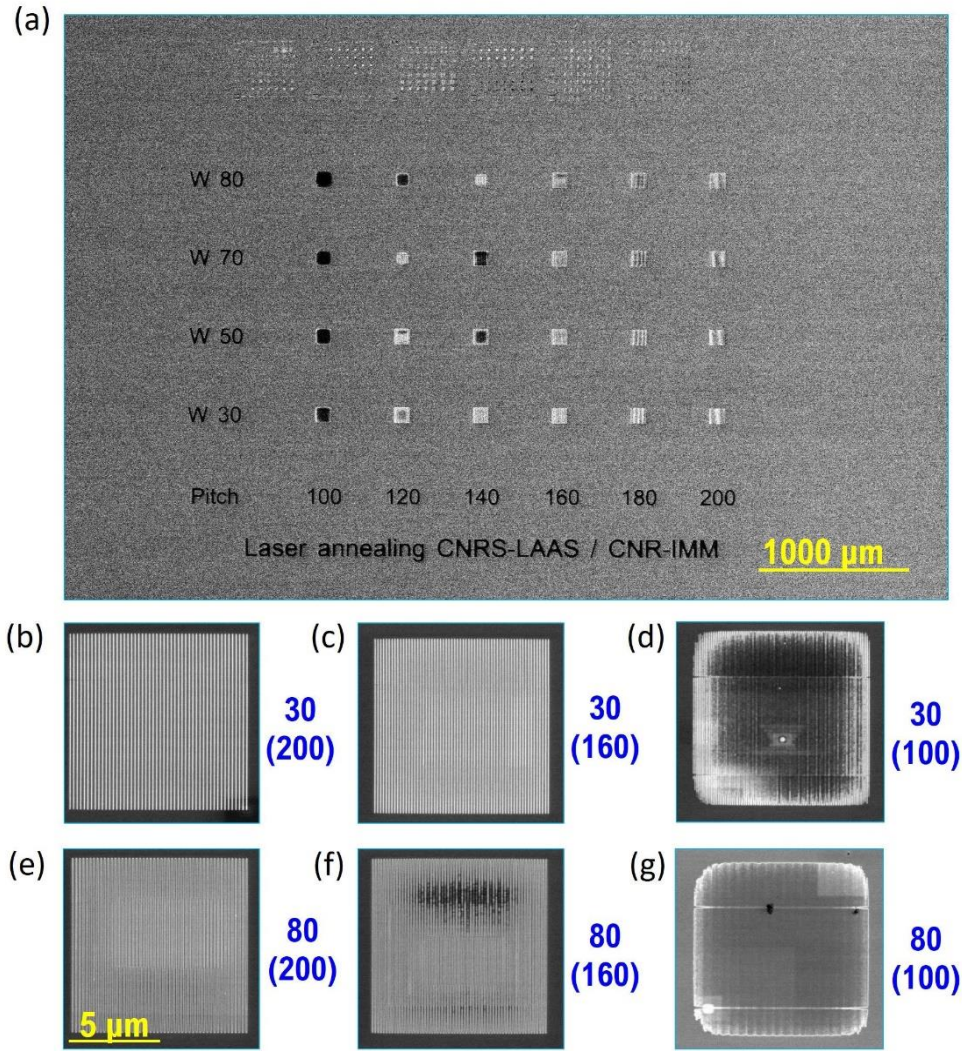

**Figure 2.** SEM micrographs of nanometer-size arrays of $SiO_2$ stripes fabricated on r-$Si_{0.5}Ge_{0.5}$ arrays prior to NLA: (a) Overview of the various arrays on each sample with nominal widths (W)/pitches (P) of (b) 30/200 nm, (c) 30/160 nm, (d) 30/100 nm, (e) 80/200 nm, (f) 80/160 nm and (g) 80/100 nm. For nominal inter-spacings between stripes of 80/70 nm in (d, f) down to 20 nm (g), enhanced proximity effects and charging during EBL result in an agglomeration (dark contrast) of the resist (HSQ), leading to a complete $SiO_2$ coverage of the surface for dense arrays as in (g).

Multiple samples such as that imaged in Figure 2 were laser-irradiated with a wavelength of 308 nm and a pulse duration ($\Delta t$) of 160 ns with different energy densities (ED) ranging from 0.20 to 0.55 J cm$^{-2}$. We selected such an ED range for the following reasons: (i) during a previous study on pure $Si_{1-x}Ge_x$ samples, i.e. without $SiO_2$ coating, we obtained an experimental (simulated) melt threshold of 0.52 J cm$^{-2}$ (0.45 J cm$^{-2}$) and (ii) we expected a slight decrease of that threshold ED due to the presence of SiO$_2$.[26] As shown in supplementary Figure S1, time-resolved reflectivity profiles recorded during higher ED laser anneals exhibited notable variations only for an energy density of 0.55 J cm$^{-2}$. This was associated with the occurrence of a solid-liquid phase transition during the process, hinting therefore at the success of the laser melting process for such an ED value. We used TRR results as a guide for the selection of annealed structures for thorough analysis with HAADF-STEM. Consequently, we conducted STEM analysis on cross-sectional lamellas of structures with nominal pitches/widths of 30/200 nm and 80/200 nm irradiated at a laser fluence of 0.55 J cm$^{-2}$. These two structures, hereafter identified as Structure 1 and Structure 2, were indeed free of coalescence phenomena during e-beam nanolithography and they represent selected cases chosen for detailed STEM analysis, to evaluate how varying the spacing between SiO$_2$ stripes impacts the laser melting process at 0.55 J cm$^{-2}$. We selected two structures with a constant P value of 200 nm to (i) streamline the investigation of the interspacing effect, now governed exclusively by the W parameter, and (ii) because, as demonstrated in Figure 2, at this P value the stripes are more distinctly resolved and completely devoid of coalescence effects. We chose upper and lower bounds for the W parameter, anticipating that W values within this range would yield effects intermediate to those analyzed.

HAADF-STEM z-contrast mode images are shown in Figure 3, with different sample magnifications. Geometrical parameters of SiO$_2$ stripes differ from their nominal values, as expected due to enhanced proximity effects during EBL on r-Si$_{0.5}$Ge$_{0.5}$. Widths are higher at the base of stripes, with narrower spacings in-between. These variations are depicted in our two-dimensional schematics presented in Figure 4a-b. In the case of Structure 1 (Structure 2), the silicon oxide layer exhibits a central flat region with a length of 26 (65) nm, slightly shorter than the nominal 30 (80) nm. This central region is enlarged on both sides by up to 17 (27.5) nm, resulting in rounded-off edges. Stripe heights are 32 nm and 40 nm, respectively. Significant differences after laser annealing are obvious for the two samples: for Structure 1, the melt depth is of 20 nm, indicating a partial-melt regime slightly above the melt threshold. Additionally, the liquid front exhibits a concave morphology with the silicon-oxide array positioned in the middle. Conversely, in the case of Structure 2, a partial-melt regime is observed with a larger melt depth of 71 nm, characterized by a rather flat liquid front. High-resolution magnifications of interfaces, provided in the bottom of Figure 3, do not show the formation of extended defects nor the presence of lateral Ge segregation, possibly due to a more confined nucleation of the liquid at the melting onset and to a homogeneous solidification process. Overall, these results indicate that the presence of differently shaped $SiO_2$ stripes had a definite impact on the efficiency of the laser annealing process.

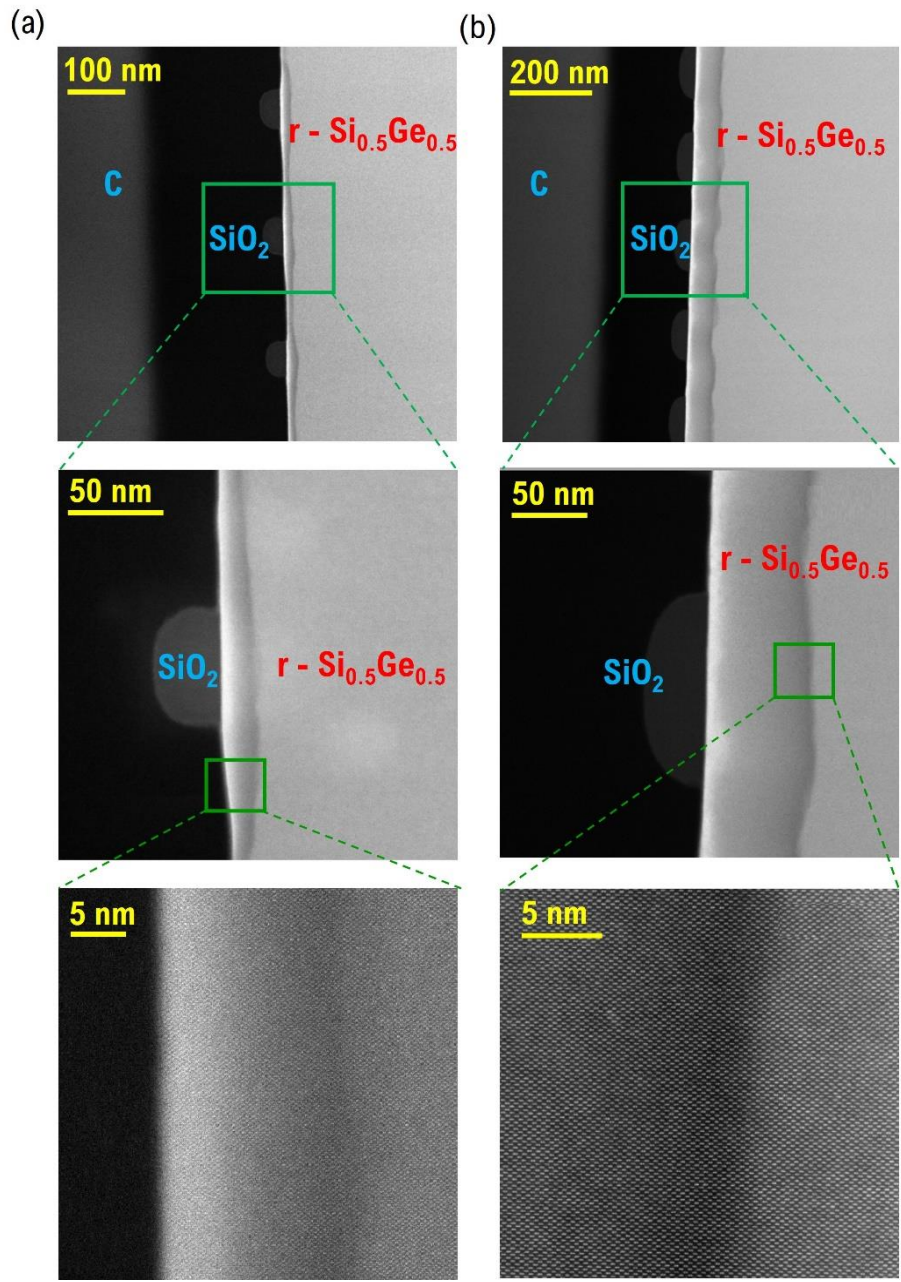

**Figure 3.** HAADF-STEM micrographs recorded in z-contrast of (a) Structure 1 and (b) Structure 2 after NLA with an energy density 0.55 J cm$^{-2}$ yielding a successful melting of the Si$_{0.5}$Ge$_{0.5}$ surface. Different magnification micrographs show the r-Si$_{0.5}$Ge$_{0.5}$ melt region and liquid front. The carbon layer used for the lamella preparation (C), SiO$_2$ stripes, r-Si$_{0.5}$Ge$_{0.5}$ and the melt depth are labelled in the pictures.

In order to obtain further information on the impact of SiO$_2$ stripes on the annealing process, we performed NLA simulations with a general methodology developed by Huet *et al* and optimized for pure Si$_{1-x}$Ge$_x$ relaxed thick samples, i.e. without the superficial silicon-oxide coating, by Ricciarelli *et al*.[26, 38] The method consists in a computationally straightforward and fast finite element method that considers the self-consistent solution of Maxwell's and Fourier's partial differential equation, where the electromagnetic field of the laser is considered as a heat source. The evolution of all key physical fields is then tracked within a two-dimensional mesh. As a general computation strategy, we simulate the evolution of the thermal field within the process utilizing the simplified enthalpy model (integrated in the general methodology) following the formalism developed by Hackenberg *et al*[39] and, when required, we simulate the Ge segregation profiles with the more extended phase field approach (also integrated in the general methodology) following the formalism by La Magna *et al.*[1] We opt for this strategy because the phase field approach is not capturing the geometric characteristics of the thermal field at the onset of melting, as it only become effective when the width of the area with the temperature surpassing the melting point exceeds ~ 8 nm. Therefore, we employ the phase field only when needed. [26, 38]

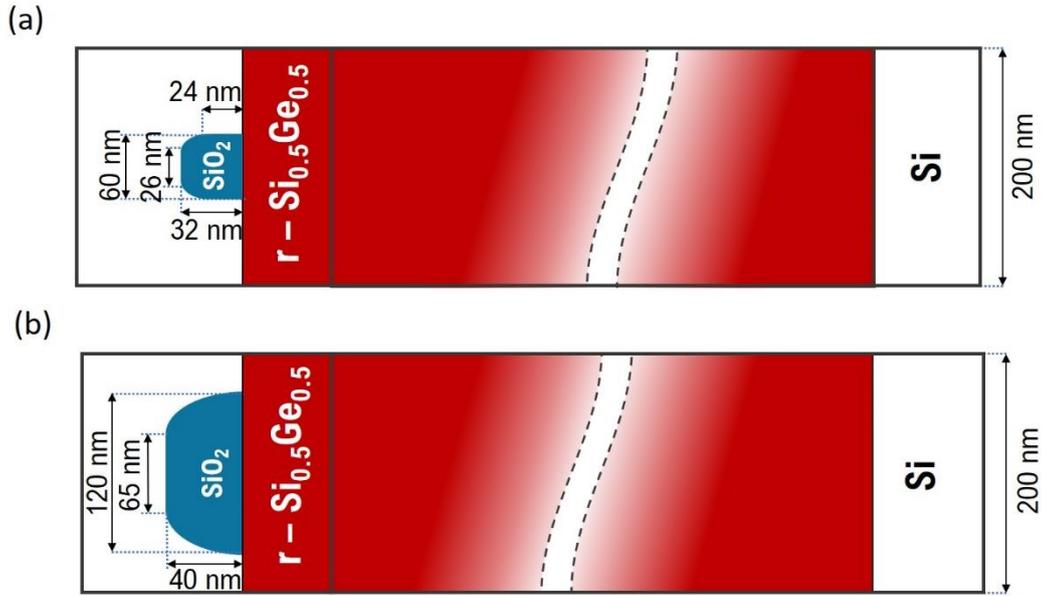

**Figure 4.** Two dimensional schemes of (a) structure 1 and (b) structure 2 based on HAADF-STEM results.

The construction of the FEM mesh was based on the schematics outlined in Figure 4 with periodic boundary conditions set along the x-axis. More specifically, we employed a 700 nm thick $Si_{1-x}Ge_x$ layer.[26] We incorporated a germanium alloy fraction of 0.50 and, additionally, a graded region of 4700 nm and 700 nm of silicon buffer were added to the bottom of the structure. At this point, we simulate the evolution of the thermal field within the process (Figure 5). Our simulations deliver a melt threshold of 0.35 J cm$^{-2}$ for Structure 1 and a slightly lower value of 0.30 J cm$^{-2}$ for Structure 2 in Figures 5a-b. This is evident from the comparison of the maximum surface temperature graphs over time. The profile flattening at the melting point means that the laser is inducing melting in the silicon-germanium substrate. These results are in line with experimental findings presented earlier. The smaller melt depth observed in Structure 1 hints to a higher melting threshold value. In both cases, the presence of the oxide greatly decreases the melt threshold, which was experimentally (computationally) determined to be 0.525 (0.45) J/cm$^2$ for pure $Si_{1-x}Ge_x$ layers.[26] Such a melt threshold lowering is due to the anti-reflection action of the oxide, in line with the increase of the oxide's width for Structure 2 compared to Structure 1. [32-36]

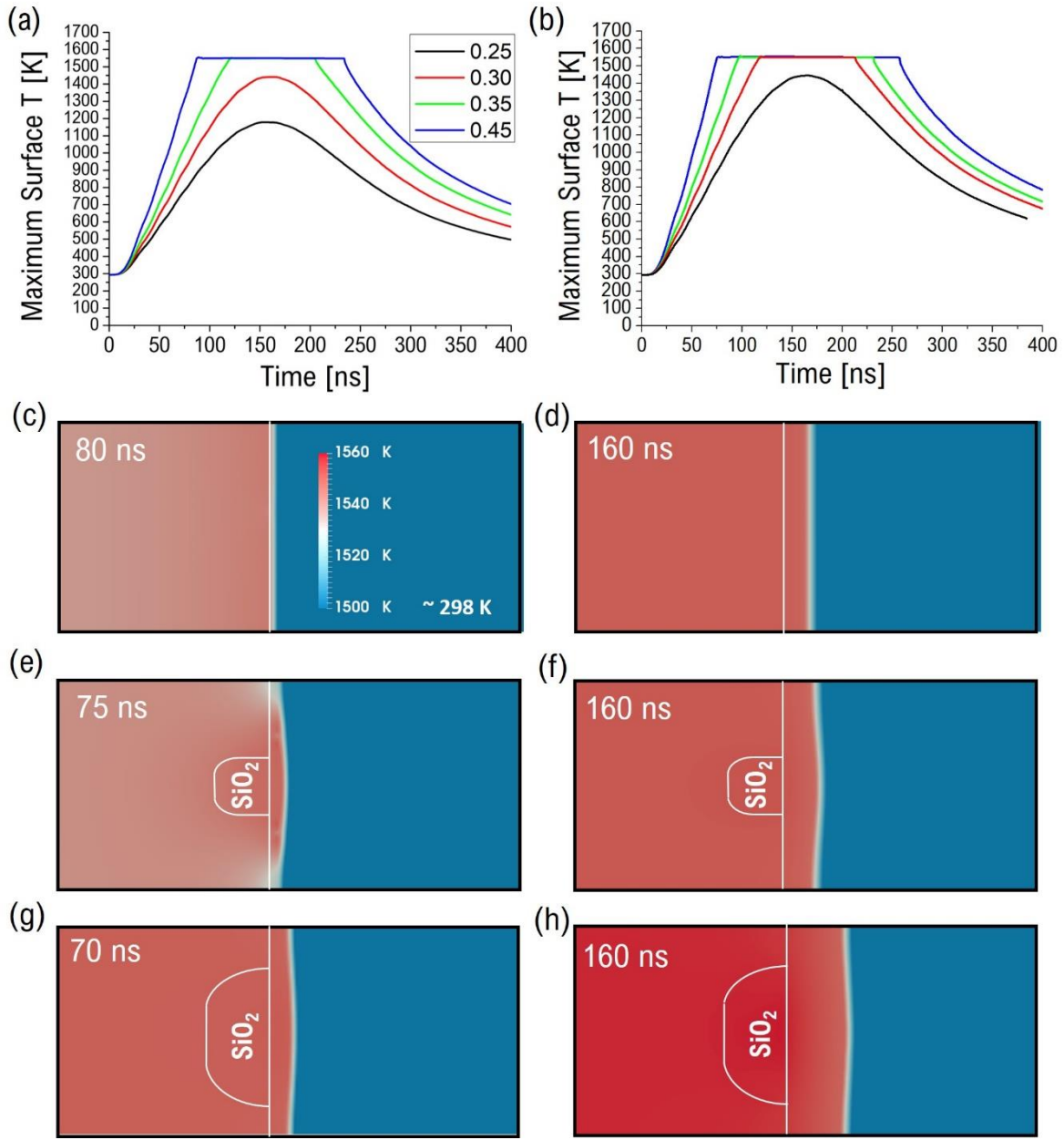

**Figure 5.** Time evolution of the maximum surface temperature for (a) Structure 1 and (b) Structure 2 for different laser energy densities (in Jcm$^{-2}$) and simulation of the different modulations of the thermal field by SiO$_2$ at ED=0.55 Jcm$^{-2}$: thermal field distribution at the melting onset for (c) pure Si$_{0.5}$Ge$_{0.5}$ Structure, (e) Structure 1 and (g) Structure 2, thermal field distribution captured at the maximum extension of the liquid front for (d) pure $Si_{0.5}Ge_{0.5}$ structure, (f) Structure 1 and (h) Structure 2. All presented data were simulated with the enthalpy model

We further emphasize the distinct modulation of the thermal field by the $SiO_2$ array influencing the distribution of the melting front, captured at the onset of melting and shown in Figure 5c-e-g. The smaller $SiO_2$ pattern in Structure 1 creates a hotspot region with temperatures equal to the alloy melting point of 1550 K (Figure 5e). The distribution of the thermal field shown in Figure 5e consequently leads to a more concave morphology of the melting front compared to Structure 2 (Figure 5g), where the temperature field appears more uniform along the x-axis. Such data explain the evolution from a concave liquid front morphology in Structure 1 (Figure 3a) to a flatter one in Structure 2 (Figure 3b). However, in Structure 2, the thermal budget is significantly higher. This is evident when comparing the geometrical distribution of the thermal field at the time of maximum liquid-state extension (Figure 5d-f-h): the field penetrates deeper inside the silicon-germanium substrate in Structure 2, with more material reaching the melting temperature than in Structure 1. Indeed, the thermal budget is at its highest for substrates fully covered by $SiO_2$, as illustrated in Figure S2. Additionally, enlargements in the H parameter within the structures also contribute to this increase (Figure S3), making it another crucial adjustable factor for optimizing thermal management.

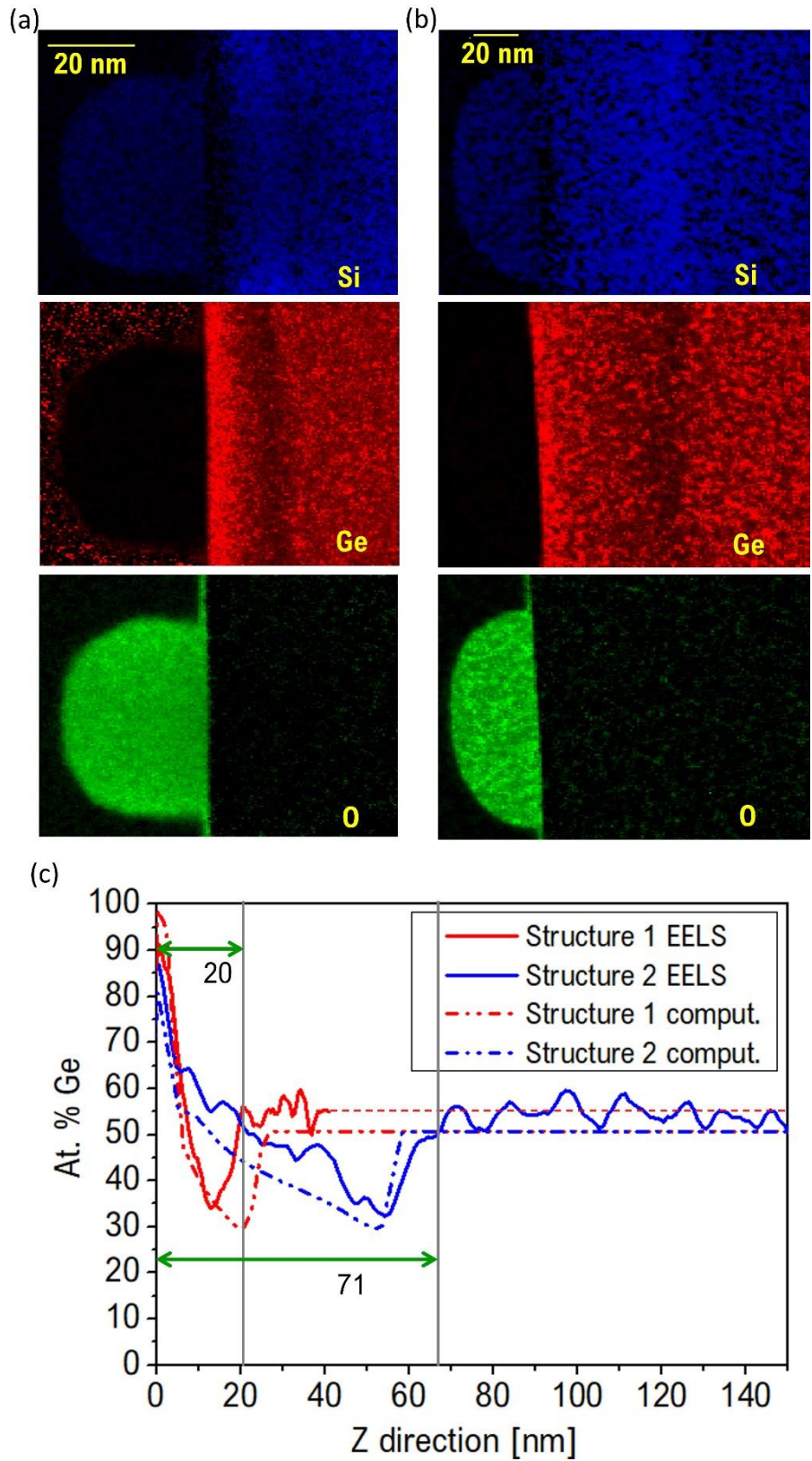

**Figure 6.** Elemental maps recorded by EELS of: (a) Structure 1 and (b) Structure 2 for silicon, germanium and oxygen (blue/red/green). (c) Comparison of germanium segregation profiles obtained from EELS and from finite element methods (comput.) modelling. Melt depths are also highlighted

by grey lines at 20 and 71 nm depth. EELS profiles are obtained by planar averaging the respective two-dimensional maps. All data are simulated with the phase field model.

Using HAADF-STEM EELS in spectrum imaging mode, we further obtain elemental maps of the structures, as shown in Figure 6a-b. The distribution of elements is uniform in the plane, while the Ge fraction fluctuates perpendicularly to the surface. This is attributed to segregation processes[27-28, 38]. Ge atoms move towards the substrate's surface, leading to superficial Ge atomic percentages of 85%-90% and creating a depleted layer at the liquid front. The Ge segregation profiles are reconstructed from the EELS L peak and shown in Figure 6c, where we added the simulated profiles (Figure 6). A closer look to those, strengthen the relationship between the silicon oxide width and the melt depth (highlighted by green arrows), the latter being larger for wider $SiO_2$ nano-arrays, due to the increased thermal budget.

**Conclusions**

In summary, we fabricate $SiO_2/Si_{1-x}Ge_x$ scaffolds with $SiO_2$ stripes atop r-$Si_{0.5}Ge_{0.5}$ ~ 700 nm thick substrates using e-beam nanolithography. We thoroughly investigate representative structures' chemical and morphological characteristics using HAADF-STEM and we employed simulations to better understand the effects of varying $SiO_2$ stripes widths on the UV-NLA process. Overall, our findings highlight that a surface coating of r-$Si_{0.5}Ge_{0.5}$ with regular arrays of nanometer scale $SiO_2$ stripes is a promising approach to enhance the control of the process during nanosecond laser annealing of $Si_{1-x}Ge_x$: investigated case were completely devoid of extended defects and lateral Ge segregation regions. Additionally, the geometrical parameters of the silicon oxide can be adjusted to fine-tune the temperature field, thereby influencing melt thresholds, melt depth, the morphology of the liquid front and the ultimate distribution of germanium. These outcomes open up avenues for the potential utilization of such nanosecond laser annealed $Si_{1-x}Ge_x$ substrates in nano-scaled devices, spanning applications ranging from microelectronics to photonics.

**Materials and Methods**

The thick relaxed $Si_{0.5}Ge_{0.5}$ layer used, referred to as r-$Si_{0.5}Ge_{0.5}$, was epitaxially grown on a 200 mm bulk Si(100) wafer (Czochralski, p-type, 1–50 Ω cm) by reduced pressure chemical vapor deposition (RPCVD) in an Epi Centura 5200C tool. $SiH_2Cl_2$ and $GeH_4$ precursors were used for that 1123K, 20 Torr deposition. Prior to SiGe epitaxy, a $H_2$ bake (at 1373 K for 2 min) was performed to remove the chemical silicon dioxide on top of Si. After surface preparation, a 4.7 µm thick linearly graded $Si_{1-x}Ge_x$ buffer was grown, with a 10% µm$^{-1}$ ramp from a few %Ge up to x=0.5. This buffer layer was capped with a 1.2 µm thick relaxed and undoped SiGe layer with a uniform Ge content of x=0.5 (same process conditions). Thanks to the high temperature (1123 K) used during the process, the glide of the threading arms of misfit dislocations (i.e. threading dislocations) was enhanced causing them to bend and remain mostly confined within the graded buffer. As a result, the threading dislocation density was significantly reduced in the $Si_{0.5}Ge_{0.5}$ top layer (~ $10^5$ cm$^2$). After RPCVD, the surface cross-hatch, e.g. the regular array of undulations along the <110> directions, was removed using a two-step chemical-mechanical polishing (CMP, planarization and smoothing) process in a Mirra CMP tool, resulting in a reduction of the top $Si_{0.5}Ge_{0.5}$ thickness from 1.2 µm down to ~0.7 µm.

Arrays of 30, 50, 70 and 80 nm wide $SiO_2$ stripes with pitches varying between 100 and 200 nm were then fabricated on r-$Si_{0.5}Ge_{0.5}$ templates by electron-beam lithography (EBL) with a Raith-150 system at minimal spot size and an acceleration voltage of 30 kV. A 50 nm thin Hydrogen Silsesquioxane (HSQ) resist layer was spin-coated on r-$Si_{0.5}Ge_{0.5}$, exposed and thereby transformed into stable $SiO_2$ forming vertical nano-lines. Unexposed resist was removed in 25% Tetramethylammonium Hydroxid (TMAH) before rinsing in Methanol. The exposure dose and write paths of lithography were adjusted depending on width/pitch combinations in order to minimize proximity and charging effects that are enhanced for Ge-rich $Si_{1-x}Ge_x$ substrates.[37]

Nanopatterned $SiO_2$/r-$Si_{0.5}Ge_{0.5}$ samples were annealed by UltraViolet Nanosecond Laser Annealing (UV-NLA) using a SCREEN LT-3100 system with a XeCl excimer laser ($\lambda$ = 308 nm, pulse duration = 160 ns, 4 Hz repetition rate, <3% laser beam uniformity, 15 x 15 mm$^2$ laser beam) at room temperature and atmospheric pressure, with a constant incident $N_2$ flux for selected energy densities (ED). In-situ Time Resolved Reflectometry (TRR) using a 635 nm wavelength laser was used to collect the light reflected by the substrate's surface before, during and after the laser pulse to detect the light intensity increase associated with the heating or melting of the sample.

TEM lamellae were prepared by focused ion beam (FIB) using a Thermofisher Helios 5 UC+ system, operating with 30keV Ga+ and finally polishing at low energy (2keV Ga+) to reduce residual FIB-induced amorphization. Scanning Electron Microscopy has been performed by using the electron column of the same system.

A JEOL ARM200F Cs-corrected microscope, equipped with a cold-field emission gun and operating at 200 keV, was used to analyze the TEM lamellae. Micrographs were acquired in Z-contrast mode by high-angle annular dark field (HAADF-STEM). A GIF Quantum ER system was used for electron energy loss spectroscopy (EELS) measurements in Spectrum Imaging (SI) mode.

Numerical simulations of the UV-NLA process, involving solid/liquid phase change and Ge diffusion phenomena were performed employing a finite element method code.[1-4, 40] The laser is considered as the heat source in the related partial differential equation and the electromagnetic field is modeled as the time-harmonic solution of Maxwell equations. The code explicitly considers the silicon-germanium liquid and solid phase with a phase-field formalism, however, to save computational sources this is only activated when the liquid layer overcomes 8 nm of critical length otherwise leaving the stage to the cheaper enthalpy model. For the article purposes, we use the phase field formalism only to calculate Ge redistribution profiles, while we use the enthalpy model in all the other cases. For more details see Refs. 3-26-38.

**Acknowledgements**

We dedicate this article in loving memory of our friend and colleague Fuccio Cristiano, who passed away in January 2024, for his extensive contribution to this work and to other related projects.

We gratefully acknowledge funding from the European Union's Horizon 2020 Research and Innovation program under grant agreement No. 871813 MUNDFAB and the European Union's NextGenerationEU under grant agreement CN00000013-National Centre for HPC, Big Data, and Quantum Computing. This work was supported by the LAAS-CNRS micro and nanotechnologies platform, a member of the Renatech french national network. The authors would like to thank SCREEN company and its French subsidiary LASSE for helping in operating and maintaining the LT-3100 laser annealing system.

**Conflicts of Interest.** There are no conflicts to declare.